\documentclass[a4paper,fleqn,usenatbib]{mnras}

\usepackage{newtxtext,newtxmath}

\usepackage[T1]{fontenc}
\usepackage{ae,aecompl}
\usepackage{threeparttable}

\usepackage{graphicx}	
\usepackage{amsmath}	
\usepackage{graphicx}
\usepackage{ulem}

\usepackage{svg}
\usepackage{amsmath}

\def\msun{M$_{\odot}$}
\def\rsun{R$_{\odot}$}

\def\it{\sl}
\def\degs{\ifmmode ^{\circ}\else$^{\circ}$\fi}
\def\amin{\ifmmode ^{\prime}\else$^{\prime}$\fi}
\def\asec{\ifmmode ^{\prime\prime}\else$^{\prime\prime}$\fi}
\def\fss{\hbox{$.\!\!^{\rm s}$}}        
\def\farcs{\hbox{$.\!\!^{\prime\prime}$}}  

\def\degs{\ifmmode ^{\circ}\else$^{\circ}$\fi}
\def\amin{\ifmmode ^{\prime}\else$^{\prime}$\fi}

\def\fermi{\textit{Fermi}}
\def\psr{J0621}
\def\psrj{J0621+2514}
\def\atnf{\url{https://www.atnf.csiro.au/people/pulsar/psrcat/}}
\newcommand{\aplerr}[3]{\ifmmode{#1^{+#3}_{-#2}} \else$#1^{+#2}_{-#3}$}

\title[Spectroscopic observation of binary MSPs]{Optical spectral  observations 
of three binary millisecond pulsars}

\author[A. Bobakov et al.]{
A. V. Bobakov,$^{1}$\thanks{E-mail: bobakovalex@gmail.com}
A. V. Karpova,$^{1}$
S. V. Zharikov,$^{2}$
A. Yu. Kirichenko,$^{2,1}$\newauthor
\ Yu. A. Shibanov,$^{1}$ and
D. A. Zyuzin$^{1}$
\\
$^{1}$ Ioffe Institute, Politekhnicheskaya 26, St. Petersburg, 194021, Russia \\
$^{2}$ Instituto de Astronom\'ia, Universidad Nacional Aut\'onoma de M\'exico, Apdo. Postal 877, Baja California, M\'exico, 22800
}

\date{Accepted XXX. Received YYY; in original form ZZZ}

\pubyear{2023}

\begin{document}
\label{firstpage}
\pagerange{\pageref{firstpage}--\pageref{lastpage}}
\maketitle

\begin{abstract}

We present the results of  optical spectroscopy of stellar companions to three binary millisecond pulsars, PSRs J0621$+$2514, J2317$+$1439 and J2302$+$4442,  obtained with the Gran Telescopio Canarias. The spectrum of the J0621$+$2514 companion shows a blue continuum and prominent Balmer absorption lines. The latter are also resolved in the spectrum of the J2317$+$1439 companion, showing that both are DA-type white dwarfs. No spectral features are detected for the J2302$+$4442 companion, however, its broadband magnitudes and the spectral shape  of the  continuum emission imply that this is also a DA-type  white dwarf. Based on the spectral analyses, we conclude that the companions of J0621$+$2514 and J2317$+$1439 are relatively hot, with  effective temperatures $T_{\rm eff}$$=$8600$\pm$200 and 9600$\pm$2000~K, respectively, while the J2302$+$4442 companion is significantly cooler,  $T_{\rm eff}$$<$6000~K. We also estimated the distance to J0621$+$2514   of 1.1$\pm$0.3 kpc and argue that its companion and the companion of J2317$+$1439 are   He-core white dwarfs providing  constraints on their cooling ages of $\la$2 Gyr. 

\end{abstract}




\section{Introduction}

Millisecond pulsars (MSPs) represent a subclass of neutron stars (NSs) with extremely short spin periods. The first MSP was discovered about 40 years ago \citep{backer1982}, and nowadays more than 550 MSPs are known\footnote{According to the ATNF catalogue \citep{atnf}; \atnf}. Most of them are found in binary systems confirming the most accepted theory that MSPs are old pulsars spun up by accretion of matter and angular momentum from donor stars during the low- and intermediate-mass X-ray binary stages \citep{Bisnovatyi-Kogan1974,alpar1982}. As a result, a system consisting of a rapidly rotating (with a spin period $P<30$ ms) pulsar and a `peeled' companion is formed. The latter often is a low-mass He-core white dwarf (WD; \citealt{tauris2011}).

Studies of binary MSPs are important since they allow one to measure NSs masses and thus to constrain the fundamental properties of supra-nuclear density matter inside them \citep{lattimer2012,antoniadis2013}. For some MSPs, this was done via measurements of post-Keplerian parameters from radio timing observations  \citep[e.g.][]{arzoumanian2018,2023MNRAS.520.1789S,2021ApJ...915L..12F,2011MNRAS.412.2763F}. For other binaries, multiwavelength, especially optical, observations are required. They provide the parameters of the companion star, such as spectral type, temperature, surface gravity, mass ratio of the system, etc., from which one can infer the pulsar mass \citep[e.g.][]{antoniadis2012,antoniadis2013,matasanchez2020}.

The optical companions of the binary MSPs  J0621+2514, J2302+4442, and J2317+1439   were recently detected at a 22--23 visual magnitude level  using photometric observations \citep{dai2017,Karpova2018K,kirichenko2018}. The authors proposed that they are likely WDs based on their magnitudes and colour indices. To confirm that and  to get additional constraints on the parameters of the companions and binary systems, we  have performed the optical spectroscopy of the objects with the Gran Telescopio Canarias (GTC). The results of the observations are presented in this paper. In Section ~\ref{sec:systems} we describe the previous studies of the targets and their parameters. Section~\ref{sec:obs} presents the observations and data reduction. The results are described and discussed in Section~\ref{sec:par} and Section~\ref{sec:conclusions}, respectively.

\section{Targets }
\label{sec:systems}

The parameters of the three binary MSPs are reported in Table~\ref{tab:param}. Below we briefly describe the previous studies of these systems.

{\bf PSR J0621+2514} (hereafter \psr) is a  $\gamma$-ray MSP discovered in the radialsoo with the Green Bank Telescope during pulsar searches among unassociated \fermi\ Large Area Telescope $\gamma$-ray sources \citep{ray2012,sanpaarsaphd}. Using the Sloan Digital Sky Survey (SDSS) and Panoramic Survey Telescope and Rapid Response System Survey (Pan-STARRS) data, \citet{Karpova2018K} identified its possible companion with a  magnitude of $\approx22$ mag in the $ g'$ band. Comparing the photometric data with WD cooling tracks they found that it is likely a He-core WD with a temperature of about 10~000$\pm$2~000 K and a mass of $\lesssim0.5$~\msun. If it has a thin hydrogen envelope, its cooling age is significantly smaller than the \psr\ characteristic age ($\lesssim 0.5$ Gyr vs 1.8 Gyr) which indicates that the pulsar can be younger. On the other hand, the  cooling  age can be consistent with the characteristic one  if the WD has a thick hydrogen envelope or its progenitor is a low metallicity star \citep{Karpova2018K}.

{\bf PSR J2302+4442} (hereafter J2302) was discovered in a search for periodic radio pulsations from \fermi\ $\gamma$-ray sources using the Nan{\c{c}}ay Radio Telescope \citep{cognard2011}. Further studies also revealed the pulsations in $\gamma$-rays. The pulsar X-ray counterpart was detected with \textit{XMM-Newton}. Using the GTC, \citet{kirichenko2018} identified the J2302 companion with the brightness $r' \approx 23.3$ mag, which is likely a He- or CO-core WD. Comparison with WD cooling tracks allowed them to estimate the companion's temperature $ 6 000^ {+1000}_{-800}$~K, a mass of $\approx 0.5$~\msun\ and a cooling age of 1--2 Gyr. Combining the estimation on the companion mass with the radio timing measurements, they also obtained constraints on the binary system inclination angle of $73_{-5}^{+6}$ degrees.

{\bf PSR J2317+1439} (hereafter J2317) was discovered in the radio with the Arecibo telescope \citep{camilo1993}. It was also detected as a $\gamma$-ray pulsar \citep{smith2017}. Its flux in the 0.1--100 GeV range is $(6.17\pm1.65)\times 10^{-13}$~erg~s$^{-1}$~cm$^{-2}$ \citep{fermi-dr3}. Optical identification of its companion with $g \approx 23$ was obtained by \citet{dai2017} with the Canada–France–Hawaii Telescope. Using the approach similar to the one described above for J0621 and J2302, they found that the companion is likely a He-core WD with a temperature of $8100\pm500$ K, a mass of $\approx 0.4$~\msun\ and an age of $\approx 11$~Gyr.

\begin{table*}
\caption{Parameters of the binary systems taken from the ATNF catalogue and \citet{sanpaarsaphd}. 
} 
\begin{tabular}{lccc}
\hline
Object                                              & J0621                             & J2302                              & J2317                               \\
\hline
Right ascension $\alpha$ (J2000)                    & 06$^{\rm h}$21$^{\rm m}$10\fss8542(1) & 23$^{\rm h}$02$^{\rm m}$46\fss978387(1) & 23$^{\rm h}$17$^{\rm m}$09\fss236381(5) \\
Declination $\delta$ (J2000)                        & +25\degs14\amin03\farcs83(3)          & +44\degs42\amin22\farcs08051(2)         & +14\degs39\amin31\farcs26102(1)         \\
Spin period $P$ (ms)                                & 2.7217879391872(4)                    & 5.19232464875420(3)                     & 3.4452510723611(5)                      \\
Period derivative $\dot{P}$ (s s$^{-1}$)            & $2.483(3)\times 10^{-20}$             & $1.3868(1) \times 10^{-20}$             & $0.24306(2) \times 10^{-20}$             \\
Reference epoch (MJD)                               & 56300                                 & 56947.00                                & 55644                                  \\
Dispersion measure DM (pc cm$^{-3}$)                & 83.629(6)                             & 13.788120(1)                            & 21.8989(2)                             \\ 

\hline

Orbital period $P_{\rm b}$ (days)                         & 1.256356677(3)                        & 125.93529692(3)                         & 2.459331465164(2)                       \\
Time of ascending node $T_{\text{asc}}$ (MJD)       & 56185.7806471(4)                      & --                                      & 55643.088185856(1)                      \\ 
Mass function $f_M$, \msun\                         & 0.001416077(9)                        & 0.009210                                & 0.002199                                \\
Minimum companion mass $M_{\rm c,min}$ (\msun)~$^a$      & 0.15                                  & 0.29                                    & 0.17                                    \\
\hline
Characteristic age $\tau_c \equiv P/2\dot{P}$ (Gyr) & 1.8                                   & 5.93                                    & 22.5                                    \\
Spin-down luminosity $\dot{E}$ (erg s$^{-1}$)       & $4.71\times10^{34}$                   & $3.9\times10^{33}$                      & $2.33\times10^{33}$                     \\
Distance $D_{\rm YMW}$ (kpc)~$^b$                        & 1.64                                  & 0.86                                    & 2.16                                    \\
Distance $D_{\rm NE2001}$ (kpc)~$^b$                     & 2.33                                  &  1.18                                   & 0.83                                    \\
Distance $D_{\rm p}$ (kpc)~$^c$                          &   --                                  & $>0.5$                                  & $2.0^{+0.4}_{-0.3}$                     \\
\hline 
Optical magnitudes~$^d$   & $u'=23(1)$,               & $r'=23.33(2)$,                      & $u=24.11(83)$, \\ 
of the companion     & $g'=21.92(6)$,                & $i'=23.08(2)$,                       & $g=22.96(5)$, \\
$band$ = value (mag) & $r'=21.76(8)$,                &                                      & $r=22.86(4)$, \\
                     & $i'=21.79(12)$                &                                      & $i=22.82(5)$             \\

\hline
\end{tabular}
\label{tab:param}
\begin{tablenotes}
\item Hereafter numbers in parentheses denote 1$\sigma$ uncertainties relating to the last significant digit quoted.  
\item $^a$~The minimum companion mass is calculated assuming the system inclination $i=90$\degs\ and the pulsar mass $M_{\rm p}=1.4$~\msun.  
\item $^b$~$D_{\rm YMW}$ and $D_{\rm NE2001}$ are the dispersion measure distances corresponding to the YMW16 \citep*{ymw} and NE2001 \citep{ne2001} models for the distribution of free electrons in the Galaxy.    
\item $^c$~$D_{\rm p}$ is the distance measured through the timing parallax (taken from \citealt{arzoumanian2018}).
\item  $^d$~Observed magnitudes of companions in different optical bands are taken from the SDSS DR17 catalogue (J0621), \citet{kirichenko2018} (J2302) and \citet{dai2017} and the SDSS DR17 catalogue (J2317).  
\end{tablenotes}
\end{table*}

\begin{figure}
    \begin{minipage}[t]{0.5\textwidth}
    \centering
        \includegraphics[trim=5.3cm 1cm 6cm 1.7cm, clip=True, width=0.7\textwidth]{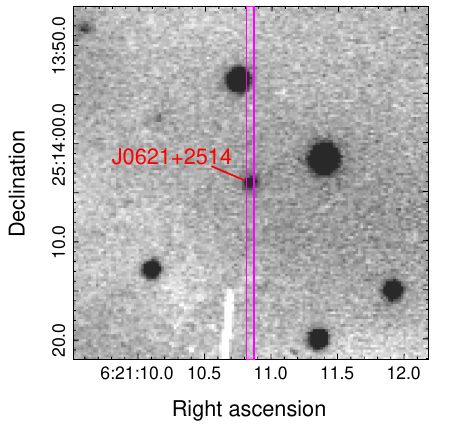}
    \end{minipage} \\
    
    \begin{minipage}[t]{0.5\textwidth}
    \centering
        \includegraphics[trim=5.3cm 1cm 6cm 1.7cm, clip=True, width=0.7\textwidth]{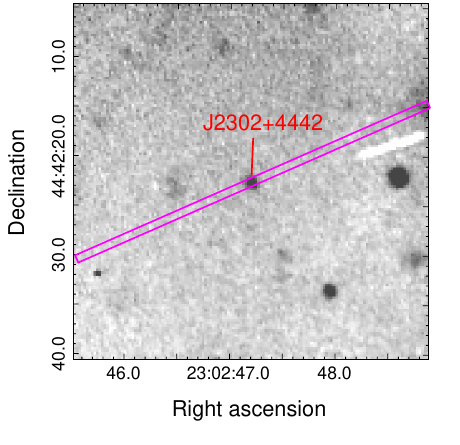}
    \end{minipage} \\
    
    \begin{minipage}[t]{0.5\textwidth}
        \centering
        \includegraphics[trim=5.3cm 1cm 6cm 1.7cm, clip=True, width=0.7\textwidth]{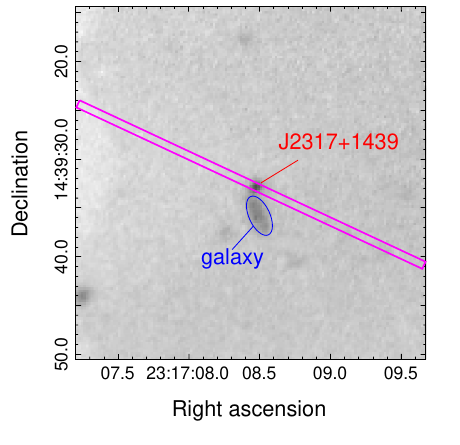}
    \end{minipage}
    \caption{0.6 $\times$ 0.6 $\text{arcmin}^2$ GTC images of the J0621 (top), J2302 (middle) and J2317 (bottom) fields obtained in the $g'$ for the top image and $r'$ for other bands. Positions of  slits with a width of 0.8 arcsec used for the spectroscopy are shown by  magenta rectangles. The optical counterparts of the pulsar companions are marked. In the bottom panel the green ellipse shows a background galaxy near the counterpart position. The  specific position of the slit was selected to minimise the contamination of the target flux by the galaxy. 
    }
\label{fig:slit}
\end{figure}


\section{Observations and data reduction}
\label{sec:obs}

\begin{table*}
	\centering
	\caption{Log of observations.
 }
	\label{tab:log}
	\begin{tabular}{lccccccccc}
		\hline
		OB & Date       & Exposure (s)  & MJD$_{\rm mid}$~$^a$ & Phase $\phi$~$^a$ &  Radial velocity$^a$ (km s$^{-1}$)    &   Airmass  & Seeing (arcsec)   & Grism   \\
		\hline
        \multicolumn{8}{c}{J0621}  \\
        1  & 2019-09-26 & 3$\times$1070 &    58766.16     &    0.75    &  $-$89$\pm$19  & 1.13--1.23 &   0.8   &        \\
        2  & 2019-10-10 & 3$\times$1070 &    58752.20     &    0.87    &  $-$326$\pm$24 & 1.12--1.21 &   1.0   & R1000B \\
        3  & 2019-11-20 & 3$\times$1070 &    58807.21     &    0.54    & -- & 1.05--1.11 &   0.9   &        \\  	
        \hline
        \multicolumn{8}{c}{J2302} \\
        1  & 2019-07-29 & 5$\times$600  &    58693.55     &     --     & -- & 1.04--1.05 &   0.9   &       \\
        2  & 2019-07-29 & 5$\times$600  &    58693.59     &     --     & -- & 1.04--1.07 &   0.9   & R300B \\
        3  & 2019-08-29 & 5$\times$600  &    58725.45     &     --     & -- & 1.04--1.07 &   0.8   &       \\
        4  & 2019-08-29 & 5$\times$600  &    58725.49     &     --     & -- & 1.04--1.05 &   0.9   &       \\
		\hline
        \multicolumn{8}{c}{J2317} \\ 
        1  & 2019-07-26 & 5$\times$545  &    58691.49     &    0.52    & -- & 1.15--1.26 &   0.9   &        \\
        2  & 2019-07-27 & 5$\times$545  &    58691.54     &    0.54    & -- & 1.05--1.09 &   0.9   & R300B  \\
        3  & 2019-07-27 & 5$\times$545  &    58691.6      &    0.56    & -- & 1.03--1.05 &   0.9   &        \\ 
        4  & 2019-08-02 & 5$\times$545  &    58697.65     &    0.03    & -- & 1.07--1.14 &   0.7   &        \\
        
        \hline
	\end{tabular}
	\begin{tablenotes}
    \item $^a$~MJD$_{\rm mid}$ is the barycentric mid-exposure time and $\phi$ is the orbital phase calculated relative to the time of ascending node using the pulsar radio ephemeris (see Table~\ref{tab:param}). Radial velocities are corrected to the Solar system barycenter.
    \end{tablenotes}
\end{table*}
\begin{figure*}
    \centering
    \includegraphics[width=\textwidth]{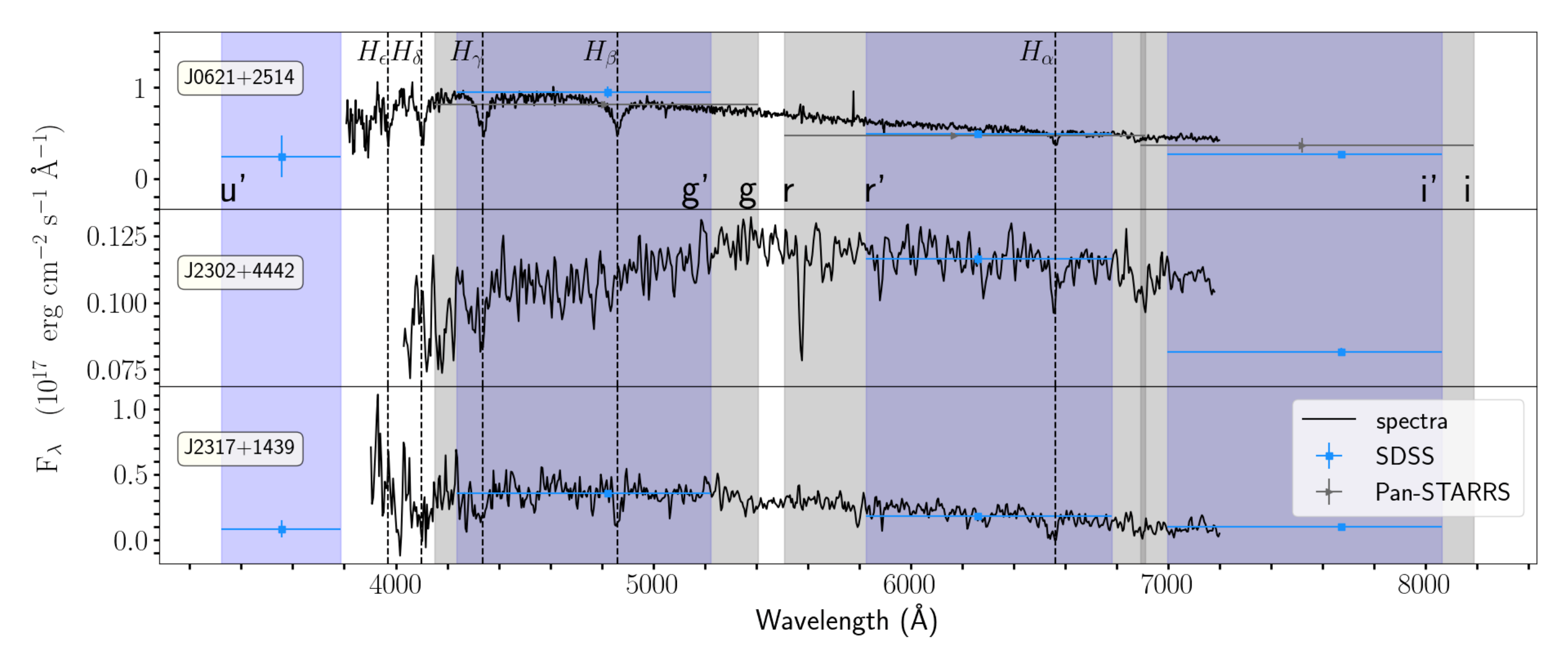}
    \caption{Calibrated spectra of the J0621, J2302 and J2317 companions. Positions of the Balmer lines are marked. Broadband photometric fluxes are shown by points with error bars as indicated in the legend. Blue strips  underline the widths of the SDSS bands,  while the light-grey strips  -- of the Pan-STARRS bands.  }
    \label{fig:calib}
\end{figure*}
\subsection{J0621}
\label{subsec:0621}

The spectroscopic long slit observations\footnote{Program GTC10-19BMEX, PI S. Zharikov} of \psr\ were performed in September and October 2019 using the Optical System for Imaging and low-intermediate Resolution Integrated Spectroscopy\footnote{\url{http://www.gtc.iac.es/instruments/osiris/}} (OSIRIS) instrument at the GTC, which at that time included two CCDs. We used the R1000B grism that covers the range 3630--7500 \AA\ and a slit width of 0.8 arcsec. The  resulting spectral resolution was 5.4 \AA. The target spectrum was exposed on the CCD2, and three observational blocks (OBs) with total integration times of 3210 s were obtained.  The slit position is shown in the top panel of Fig.~\ref{fig:slit} and the log of the observations is presented in Table~\ref{tab:log}.

The standard data reduction was performed, including the bias subtraction and flat-fielding, using the {\sc gtcmos} tool \citep{gtcmos} based on the Image Reduction and Analysis Facility ({\sc iraf}) package. We extracted the companion 1D spectra from the 2D images with the {\sc apall} procedure.

We performed the wavelength calibration using the HgAr and Ne lamp frames. Positions of each emission line of the lamps were measured and then the resulting values were fitted with a cubic polynomial to calculate the dispersion relation along the dispersion direction. This calibration was stable during the different nights of observations resulting in the root-mean-square (rms) residuals scattered closely to $0.045$ \AA which corresponds to $2.5$ km~s$^{-1}$ at 5455 \AA.

To implement the flux calibration, the spectrophotometric standards {\it Ross 640} (OB 1) from \citet{Oke1974-std-ob1}, {\it G24-9} (OB 2) from \citet{Oke1990-std-ob2} and {\it G158-100} (OB 3) from \citet{Oke1990-std-ob2} were observed during the same nights as the target, with a slit width of 2.52 arcsec. The standards were exposed on the same CCD position as the target. The spectra of the standards were reduced in a similar way as described above. Using their 1D spectra, and the atmospheric extinction curve provided by the GTC team,\footnote{\url{https://www.ing.iac.es//Astronomy/observing/manuals/ps/tech_notes/tn031.pdf}} we found the CCD response functions.  The latter were applied  to the target spectra for the preliminary flux calibration. As the slit width used for the target  was significantly narrower than that for the standards, we additionally corrected the  derived spectral flux densities of the target for the light losses at  the narrow slit. To do that, we calculated the fluxes in the broad bands from the SDSS DR17 \citep{sdssdr17} and Pan-STARRS DR2 \citep{Pan_starrs_database_Flewelling_2020} catalogues which overlap with the observed spectral range. We used  the preliminary flux calibrated target spectra obtained above and the transmission curves  of the bands. Then we fitted the calculated fluxes with  the tabulated ones (Table~\ref{tab:param}) using  an additive constant as the single fit parameter.

As a result, we obtained the calibrated spectra of the companion with the continuum signal-to-noise ratio (SNR) of $\sim 10$ for the first and second OBs and $\sim 3$ for the third OB.  The low SNR in the third OB is caused by the weather conditions such as clouds and moon illumination, so we excluded it from further consideration. To increase the  SNR, we shifted the spectra in OB1 and  OB2 to the zero velocity position using the radial velocity estimates $V_{\rm OB1} = -89 \pm 19$ km~s$^{-1}$ and $V_{\rm OB2} = -326 \pm 24 $ km~s$^{-1}$ derived from the spectral line cross-correlation analysis. We then combined the spectra to get the averaged one. This resulted in the continuum SNR $\sim 15$. The spectrum and the photometric data points are shown in Fig.~\ref{fig:calib}.


\subsection{J2302 and J2317}
\label{subsec:2302-2317}
J2302 and J2317 were observed\footnote{Program GTC2-19AMEX, PI A. Kirichenko} in summer 2019 with the OSIRIS instrument using the R300B grism, in the wavelength range  3600--7200 \AA\ and with a slit width of 0.8 arcsec. The resulting spectral resolution was 15 \AA\ at 5400 \AA. As compared to J0621, the low resolution was selected to get a reasonable SNR since these two objects are by a magnitude fainter  than  J0621 (cf. Table~\ref{tab:param}). The target spectra were exposed on CCD2. The slit positions are presented in the middle and bottom panels of Fig.~\ref{fig:slit} and the log of observations -- in Table~\ref{tab:log}. For each of the objects, the observations  were distributed over four OBs with five exposures of about 10 min. The sky was clear during all observing runs.

The standard data reduction, wavelength, and flux calibration were performed in the same way as for J0621. We used HgAr and Ne calibration lamp frames for the wavelength calibration and {\it Feige 110} \citep{Oke1990-std-ob2} spectrophotometric standard for the flux calibration. SNRs of calibrated spectra for various OBs varied from 2.5 to  3.0 and   from 6.5 to 7 for J2302 and  J2317 respectively. 

Low spectral resolution and SNRs, and poor line profiles  in the case of  J2302 and J2317  do  not allow us to confidently measure  the radial velocities at different OBs. However, the difference between the velocities can be neglected  as their  orbital periods are significantly longer than  the duration of observations (cf. Tables~\ref{tab:param} and \ref{tab:log}). To increase the SNRs, we thus simply combined the spectra of the respective OBs and obtained averaged spectra with the resulting SNRs $\sim 5 $ and $\sim 11$ for J2302 and  J2317, respectively.

The  spectra and the photometric  points are shown in the middle and bottom panels of Fig.~\ref{fig:calib}.

\begin{figure*}
    \centering
    \includegraphics[width=\textwidth]{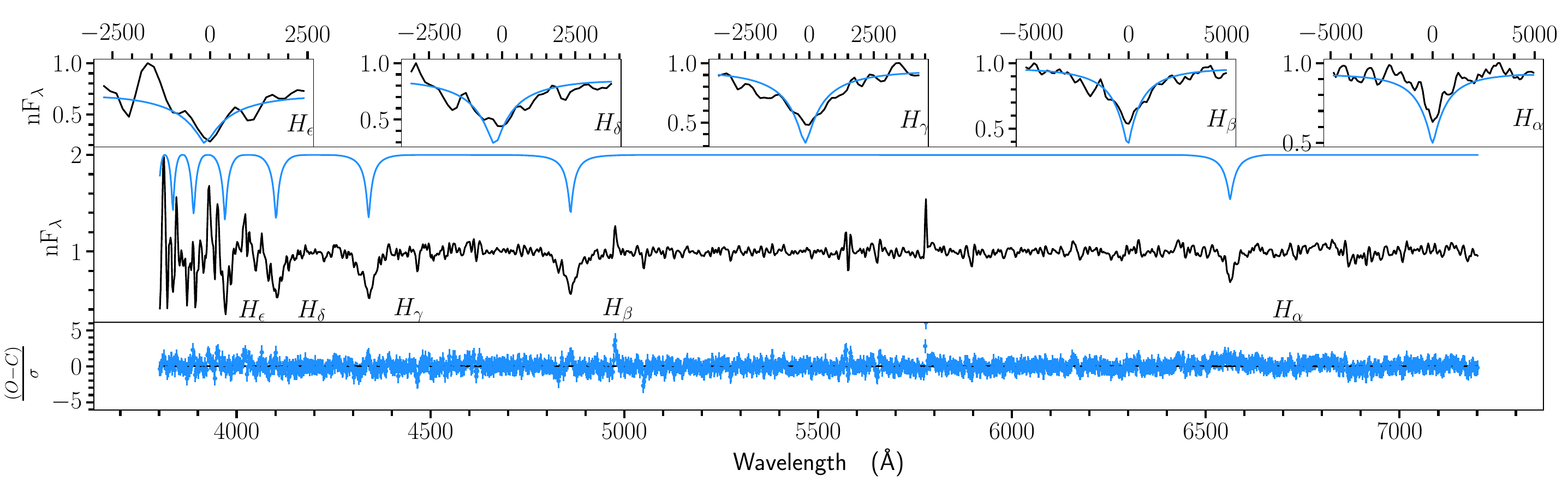}
    \caption{Normalised spectrum $nF_{\nu}$ of the J0621 companion shown by  the black line in the middle panel. The blue line is the best-fitting model ($\log(g) = 6.5$, $T_{\rm eff} = 8600$~K) shifted upwards by 1.0. The fit residuals between the observed (O) and the calculated model (C) spectra in 1$\sigma$ units are presented in the bottom panel. The regions of Balmer lines with the best-fitting model overlaid are  in the top panels where the horizontal   axis is the radial velocity units of km s$^{-1}$.}
    \label{fig:norm_J0621}
\end{figure*}
 \begin{figure*}
     \centering
     \includegraphics[width=\textwidth]{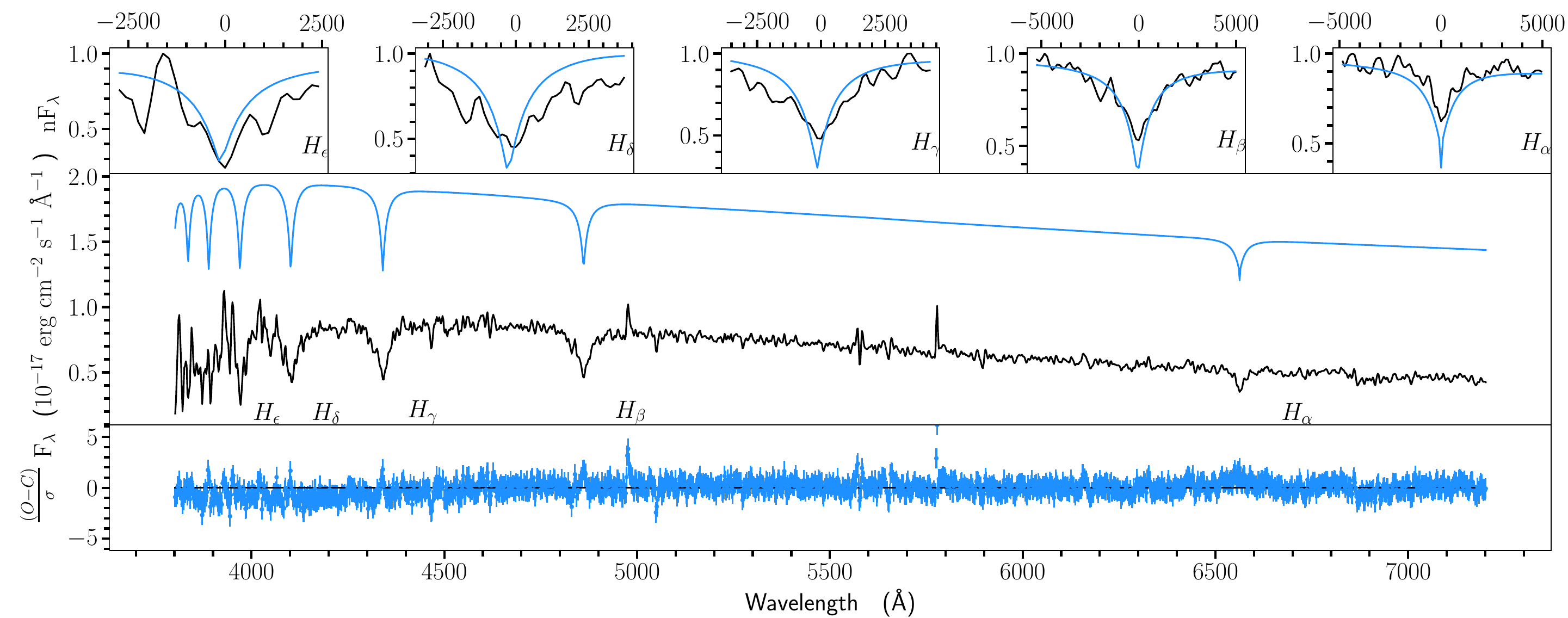}
     \caption {The same as in Fig.~\ref{fig:norm_J0621}, but in the case of the non-normalised spectrum $F_{\lambda}$. The best-fitting model in the middle panel is shifted up by $10^{-17}$ erg~cm$^{-2}$~s$^{-1}$~\AA$^{-1}$ for clarity.  The best-fitting distance is $D = 1100$ pc and $E(B-V)=0.19$. }
     \label{0621-final-fit}
 \end{figure*}

\section{Spectral analysis and parameters of the companions}
\label{sec:par}

As seen from Fig.~\ref{fig:calib}, Balmer absorption lines are prominent in the spectrum of the J0621 companion and marginally resolved for J2317, indicating that they are DA WDs. No lines are seen for J2302, likely due to a low effective temperature. We estimated the parameters of the companions  by fitting the resulting spectra with the theoretical spectral models of DA WD atmospheres calculated for a grid of values of the effective temperature $T_{\rm eff}$ and the surface gravity $g$   \citep{koester2008}. To do this, we expressed the model spectral flux density as  $F_{\lambda}^m = \pi(R / D)^2 \times  f_{\lambda}^m$,  where $R$ is the companion radius, $D$ is the distance to the source, and $f_{\lambda}^m$ is the  tabulated atmospheric flux calculated by \citet{koester2010}\footnote{The models can be found at the Theoretical spectra web server \url{http://svo2.cab.inta-csic.es/theory/newov2/}} for $ 5500~K \leqslant T_{\rm eff} \leqslant 80000~K$ and $ 6.5 \leqslant \log(g) \leqslant 9.5$. A linear interpolation was applied to get $f_{\lambda}^m$ between the $T_{\rm eff}$--$g$ grid nodes. The model spectra were convolved with a Gaussian kernel of 5 \AA ~ to decrease their spectral resolution to that of the observed spectra. We also included in the model the interstellar extinction using  the extinction--distance relations along the line of sight for each of the objects taken from the 3D dust Galaxy map of \citet{green2019}. The distance in the relations was linked with that in the expression for $F_{\lambda}^m$ presented above, and the extinction curve model from \cite{fitzpatrick2019} was applied. The fitting parameters were $T_{\rm eff}$, log$(g)$, $R$, $D$ and the interstellar reddening $E(B-V)$.     
\begin{figure*}
    \centering
    \includegraphics[width=\textwidth]{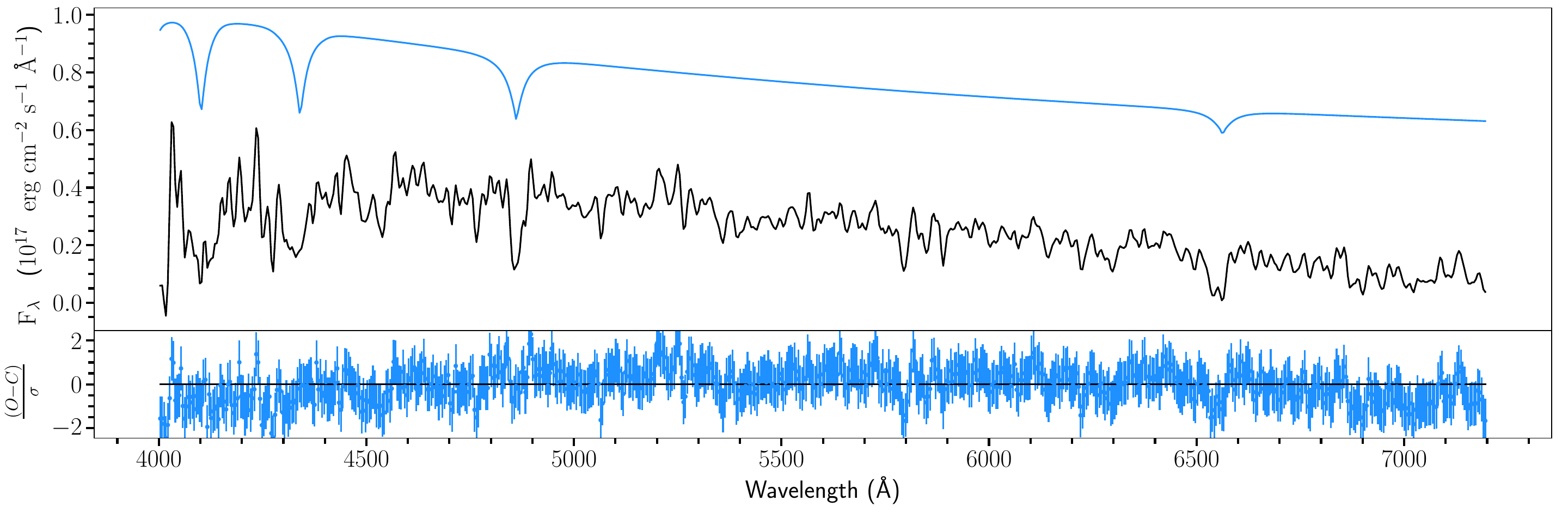}
    \caption{Observed spectrum of the J2317 companion and the best-fitting model with  fixed  $D=830$ pc shown
     by the black and blue lines, respectively, in the top panel. The best-fitting model is shifted up by $0.5\times 10^{-17}$ erg cm$^{-2}$ s$^{-1}$ \AA$^{-1}$ for clarity. 
     Fit residuals between the observed (O) and the calculated model (C) spectra in 1$\sigma$ units are presented in the bottom panel. }
    \label{fig:J2317_fit}
\end{figure*}
\begin{figure*}
    \centering
    \includegraphics[width=1.0\textwidth]{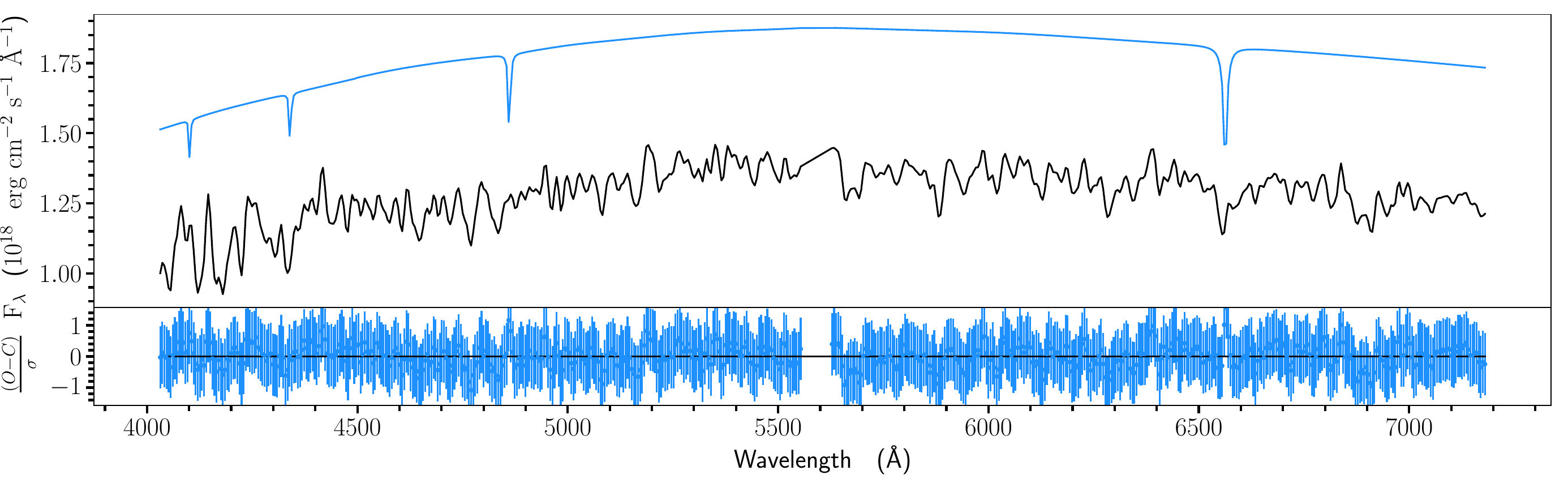}
    \caption{Observed  spectrum of the J2302 companion and the best-fitting model with fixed $D=900$ pc shown
     by the black and blue lines, respectively, in the top panel. The best-fitting model is shifted up by $0.5\times 10^{-18}$ erg cm$^{-2}$ s$^{-1}$ \AA$^{-1}$ for clarity.
     Fit residuals between the observed (O) and the calculated model (C) spectra in 1$\sigma$ units  are presented in the bottom panel.  }
    \label{fig:J2302_fit}
\end{figure*}

\subsection{J0621}

At an initial step of finding the spectral parameters,  we normalised the model and the observed spectra using a low-order polynomial continuum fit. Normalised spectra  depend mostly on $\log(g)$ and $T_{\rm eff}$ parameters, and we derived the latter from the spectral fit. This resulted in $\log(g) = 6.5 \pm 0.8$, $T_{\rm eff} = 8600 \pm 200 $~K at $\chi^2 = 0.41$ per 1379 degrees of freedom (dof). The best-fitting results are shown in Fig.~\ref{fig:norm_J0621}. The derived $T_{\rm eff}$ and $\log(g)$ are in a good agreement with the estimates obtained  using the photometric data \citep{Karpova2018K}. Their large uncertainties  are related to the low SNRs within  the observed spectral lines.   

At the next step, given $g$ and $T_{\text{eff}}$, we estimated the companion radius $R_{\text{J0621}}$ and mass $M_{\text{J0621}}$ using the relativistic mass-radius relation for DA WDs \citep{Mass-rad_rel_WD} and the surface gravity definition $g=GM/R^2$, where $G$ is the gravitational constant. This leads  to $R_{\text{J0621}}=13^{+9}_{-6}\times10^3 ~ \text{km}$  or $0.028^{+0.010}_{-0.012}$ \rsun.  The mass is very uncertain, lying in the range of 0.03--0.26~\msun. Formally, the best-fitting $M_{\text{J0621}}= 0.09$ \msun.              

 Finally, we returned to the non-normalised spectra. Fixing $T_{\rm eff}$, $\log(g)$, and $R_{\text{J0621}}$ at the values obtained at previous steps, we fitted the observed spectrum using only the distance as a free parameter and accounting for the reddening based on the distance-extinction relation. The resulting distance  is $D =  1100 \pm 300$ pc  and $E(B-V)=0.19\pm0.02$.  The final fit results are demonstrated in Fig.~\ref{0621-final-fit} ($\chi^2=0.4 ~ \text{dof}=1380$ ) and the best-fitting parameters are  presented in Table~\ref{tab:J2302_J2317_fit_pars}. 
 
\begin{table*}
    \centering
    \centering
    \caption{ Parameters of  pulsars'  companions derived from spectral data. 
    } 
    \begin{tabular}{c|ccccc}
        \hline
         $D ~(\text{kpc})$ & $E(B-V)$ & $T_{\text{eff}} ~(\text{K})$ & $\log(g)$ & $M_{\rm c}$~(\msun) & $R_{\rm c}$~(\rsun)\\     
        \hline
        \multicolumn{6}{c}{J0621} \\
           1.1$\pm$0.3 & $0.19 \pm 0.02$   & $8600 \pm 200 $& $ 6.5 \pm 0.8$ & $0.09^{+0.26}_{-0.06}$ &  $0.028^{+0.010}_{-0.012}$\\ 
        \hline
        \multicolumn{6}{c}{J2302 $^a$} \\
           0.5 & $0.12 \pm 0.04$ & $ < 5500$ & $ 8.1 \pm 1.0$ &  $0.64^{+0.60}_{-0.44}$ &  $0.012^{+0.009}_{-0.006}$\\
           0.9 & $0.13 \pm 0.02$ & $ < 5500$ & $ 6.8 \pm 1.0$ & $0.16^{+0.37}_{-0.12}$ &  $0.024^{+0.020}_{-0.010}$\\
           1.2 & $0.18 \pm 0.02$ & $ < 6000$ & $ < 6.5$ & $< 0.09$ &  $ > 0.028$\\
		\hline
        \multicolumn{6}{c}{J2317 $^a$} \\ 
            0.8 & $0.05 \pm 0.02$ & $10500 \pm 2000$ & $8.6 \pm 1$   & $0.53^{+0.60}_{-0.37}$ &  $0.012^{+0.009}_{-0.006}$ \\
            2.0 & $0.05 \pm 0.02$ & $ 9600 \pm 2000$ & $7.0 \pm 1$ & $0.14^{+0.33}_{-0.10}$ &  $0.022^{+0.014}_{-0.009}$\\
            2.2 & $0.05 \pm 0.02$ & $ 9400 \pm 2000$ & $6.8 \pm 1$ & $0.11^{+0.27}_{-0.08}$ &  $0.025^{+0.015}_{-0.010}$\\
\hline        
        
    \end{tabular}
   
    \begin{tablenotes} 
        \item $^a$~Parameters are derived at distances fixed at their three possible values (see Table~\ref{tab:param}). 
    \end{tablenotes}
    \label{tab:J2302_J2317_fit_pars}
\end{table*}

\subsection{J2302 and J2317}

As seen from Fig.~\ref{fig:calib}, the Balmer lines are weak and  poorly resolved for both objects. In contrast to the J0621 case, this makes the normalised spectra  useless for convincing estimates of $T_{\rm eff}$ and $g$. Therefore, to fit the spectra of these objects, we fixed $D$ in the $(R/D)^2$ factor of the model at the values provided by the radio data presented in Table~\ref{tab:param}. $E(B-V)$ was taken from the extinction-distance relations. To get the rest of the parameters, $R$ was linked with  $\log(g)$  through the  mass--radius relation.  As a result, for each of the objects, we found  temperatures and surface gravities for their three  possible distance values (Table \ref{tab:J2302_J2317_fit_pars}). Examples of the best-fitting results are demonstrated in  Fig.~\ref{fig:J2317_fit} and \ref{fig:J2302_fit}. Using the fitting results we can constrain the masses of the WDs. For the J2317 distance range, the mass $M_{\rm J2317}$~=~0.12--0.55~\msun, and for J2302 the mass is $M_{\rm J2302} < 0.59$~\msun. For J2302 the resulting temperature reaches the lower bound of the available atmospheric models, and thus we set only an upper limit for this object. 

\begin{figure}
\begin{minipage}[h]{1.\linewidth}
\center{\includegraphics[width=1.\linewidth,clip]{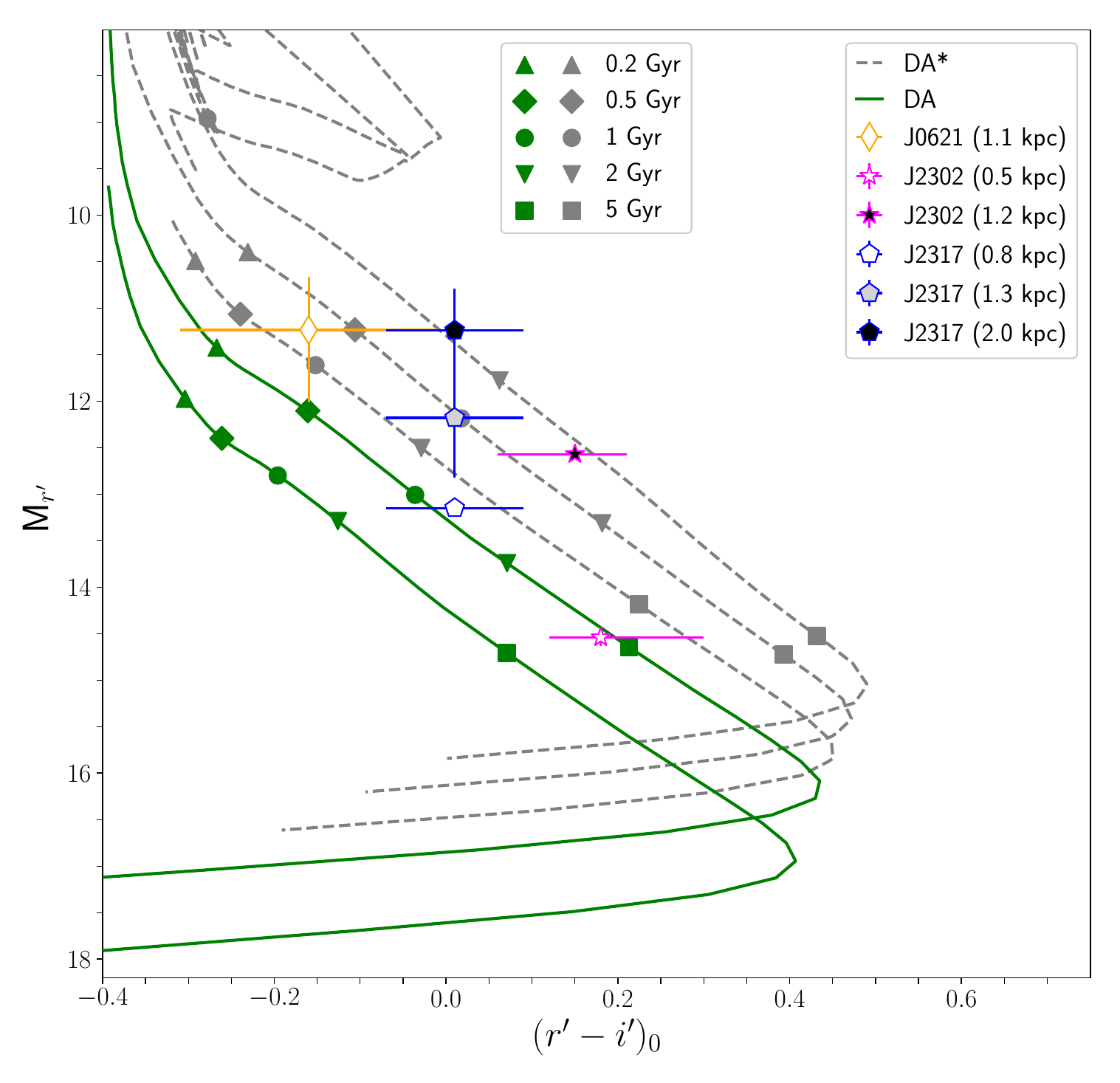}}
\end{minipage}
\caption{Colour-magnitude diagram with theoretical WD cooling tracks. Dashed lines labelled as DA$^*$ correspond to the models for He-core WDs with hydrogen atmospheres and masses 0.1821, 0.2724 and 0.4352 \msun\ \citep{althaus2013}, while the solid green lines labelled as DA -- for CO-core WDs with hydrogen atmospheres and masses 0.6 and 1 \msun\ \citep{holberg2006,kowalski2006,tremblay2011}. Masses increase from upper to lower tracks. Positions of the J0621, J2302 and J2317 companions for different distance estimates are shown by various symbols as indicated in the legend. WD cooling ages are also marked. 
}

\label{fig:cool}
\end{figure}

\section{Discussion and conclusions}
\label{sec:conclusions}

Balmer absorption lines detected in the spectra of J0621 and J2317 companions clearly show that they are DA WDs. The  spectrum of the J2302 companion  is apparently featureless though it also can be described by the coldest available hydrogen atmosphere model, and thus the companion likely belongs to the DA family  as well. 

The companions of J0621 and J2317 are relatively hot, with $T_{\rm eff}$ in the ranges 8400--8800 and 7400--12500~K, respectively, while the J2302 companion is significantly cooler, with $T_{\rm eff} < 6000$~K (see Table \ref{tab:J2302_J2317_fit_pars}). The temperature of J0621 is in agreement with the rough estimates obtained by \citet{Karpova2018K} using photometry data alone, while its uncertainty is now  much smaller. The spectral data for the J2302 companion  completely exclude temperatures above 6000~K which were allowed by the estimates based only on its photometry   \citep{kirichenko2018}. The spectral temperature of J2317 is consistent with the photometric one \citep{dai2017} while its uncertainty remains pretty large.       

The better quality of the J0621 companion spectrum enabled us to independently estimate the distance to the system of 1.1$\pm0.3$ kpc. At $\sim$2$\sigma$ level it is consistent with the dispersion measure distance of   $\approx$1.6 kpc based on the NE2001 \citep{ne2001} model for the distribution of free electrons in the Galaxy and it disagrees with that of   $\approx$2.3 kpc following from the YMW16 distribution \citep*{ymw}. Using this new spectroscopic distance, the respective $E(B-V)$ from Table~\ref{tab:J2302_J2317_fit_pars}, and the stellar magnitudes from Table~\ref{tab:param}, we calculated  an updated companion position at the dereddened colour-magnitude diagram (CMD) presented in Fig.~\ref{fig:cool} \citep[c.f.,][]{Karpova2018K}.  To understand its core composition and evolution stage, we also show the cooling tracks for DA WDs with He-\footnote{\url{http://evolgroup.fcaglp.unlp.edu.ar/TRACKS/tracks_heliumcore.html}} \citep{althaus2013} and CO-cores\footnote{\url{http://astro.umontreal.ca/~bergeron/coolingModels/}} \citep{holberg2006,kowalski2006,tremblay2011}. As seen from the CMD, the position of  J0621 is consistent with the tracks of He-core WDs  for the mass range from 0.16 to 0.44~\msun. The mass of this companion is poorly constrained from the spectral data while its upper bound of 0.35~\msun\ (see Table~\ref{tab:J2302_J2317_fit_pars}) is compatible with the above mass range supporting the He-core WD origin. From the coincidence with the tracks, the corresponding WD cooling age $t_{\rm cool}$ is $\lesssim 2$~Gyr.  

The CMD positions of the rest of the companions remain very uncertain mainly due to the distance uncertainties. As seen from Fig.~\ref{fig:cool}, they can be either He or CO core WDs depending on the distance. 

Nevertheless, for the J2317 companion, the $P_{\text{b}}$--$M_{\rm c}$ relation \citep{tauris1999} predicts the mass $\approx 0.2$~\msun, which lies within the mass range derived from the spectroscopy (Table~\ref{tab:J2302_J2317_fit_pars}). This value is in an excellent agreement with the J2317 position  in the CMD  (Fig.~\ref{fig:cool}) at the distance upper bound of  $\approx 2$~kpc when it lies at the respective DA$^*$  cooling track. The WD cooling age, in this case, is $\lesssim 2$~Gyr. A twice larger mass estimate obtained by \citet{dai2017} from the photometry for $D=1.3^{+0.4}_{-0.3}$ kpc is incompatible with the theoretical predictions indicating that the larger distance may be more reasonable. We note that even for this $D$, the conservative constraint on the WD cooling age is $0.5\ {\rm Gyr} \lesssim t_{\rm cool}\lesssim 2.5$~Gyr, while \citet{dai2017} provided $t_{\rm cool}\approx 11$~Gyr. The latter seems to be unreasonably large for a WD with $T_{\rm eff}\sim 10 000$~K. This discrepancy is likely related to the fact that the ages in the models of \citet{althaus2013} are counted from the zero-age main sequence and certain corrections are needed to obtain the real cooling ages.

Large uncertainty of the J2302 companion mass  obtained from the spectroscopy does not allow us to derive its core chemical composition. At this stage, both the He- and CO-core are allowed for this WD. 


\begin{figure}
    \centering
    \includegraphics[width=\linewidth]{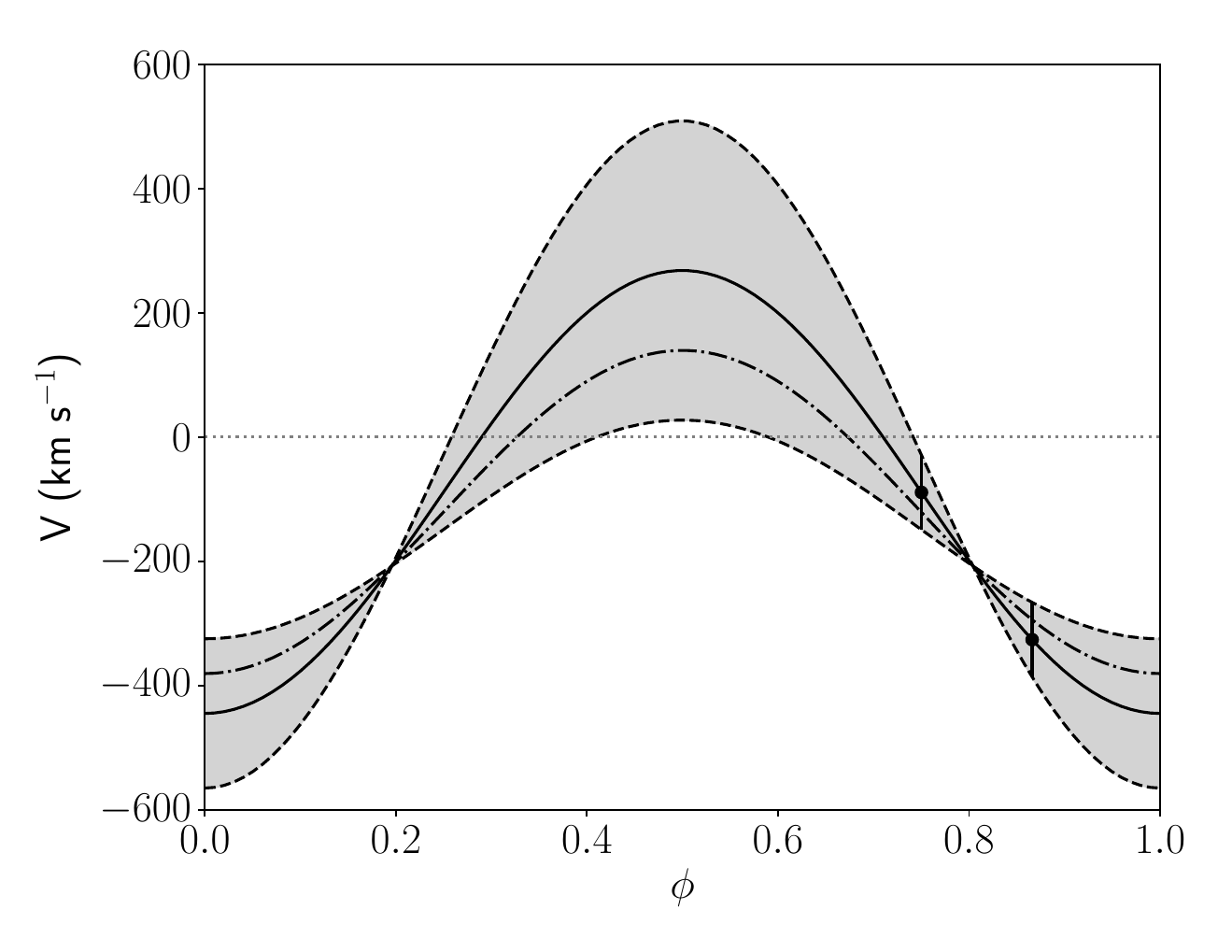}
    \caption{Radial velocity measurements for J0621 companion (black dots with 2$\sigma$ error bars). The black solid line corresponds to the best fit of the data by the model of the radial velocity curve. Black dashed lines show the $2\sigma$ boundary of the fit uncertainty (the shadowed region).    The dashed-dotted line correspond to the radial velocity semi-amplitude of 260 km~s$^{-1}$ which is the maximal value based on  the mass function given by the radio data and the WD mass range provided by the optical spectroscopy.}
    \label{fig:rvcurve}
\end{figure}

It is instructive  to consider what information can be extracted from the two radial velocity measurements of J0621 taken at its different orbital phases (see Table~\ref{tab:log}). To estimate   possible ranges of the radial velocity semi-amplitude of the companion $K_{\rm c}$ and the systemic velocity  of the system $\gamma$, we fitted the measurements with the model of the radial velocity curve
\begin{equation}
   V = K_{\rm c} {\rm cos}(2\pi\phi) + \gamma, 
\end{equation}
where $\phi$ is the orbital phase. The results are presented in Fig.~\ref{fig:rvcurve}. The derived 2$\sigma$ uncertainty ranges for $K_{\rm c}$ and  $\gamma$  are 175--536 km~s$^{-1}$ and  27--148 km~s$^{-1}$,  respectively. Such large  uncertainties are mainly caused by the presence of only  two observational points. Nevertheless, the $\gamma$ range  is in agreement with the observed velocity distribution for binary pulsars \citep{hobbs2005}. Given the binary period  $P_{\text{b}}$ and the  mass-function $f_{\text{M} }$ provided by the radio observations  (Table ~\ref{tab:param}), the pulsar radial velocity semi-amplitude $K_{\text{p}}$ can be estimated  as 
\begin{equation}
K_{\text{p}} = P_{\text{b}}\sqrt[3]{2 \pi G f_{\text{M} }} = 22~{\rm km~s}^{-1},    
\end{equation}
where $G$ is the gravitational constant. This results in the mass ratio $M_{\rm p}/M_{\rm c} \equiv K_{\rm c}/K_{\rm p}$=7.9--24.2. 

Alternatively, $K_{\text{c}}$ can be expressed trough $K_{\text{p}}$, $f_{\text{M}}$, and $M_{\rm c}$   
\begin{equation}
K_{\text{c}} = K_{\text{p}} \sqrt{M_{\text{c}}~\sin^3(i)~f^{-1}_{\text{M}} - 1},    
\end{equation} 
where $i$ is the orbital inclination. Using $M_{\rm c}$ from our  spectroscopy (Table~\ref{tab:J2302_J2317_fit_pars}), this results in $K_{\text{c}} \le 260$ km~s$^{-1}$ at $i\le90$ deg. The radial velocity curve corresponding to this upper bound is shown by the dash-dotted line in Fig.~\ref{fig:rvcurve}.  As can be seen, this line is compatible with the radial velocity fit within 2$\sigma$ uncertainties. However, the corresponding upper bound of the mass ratio is 11.7,  about twice lower than that obtained  from the fit. For our $M_{\text{c}}$  estimates, higher values of the mass ratio are inconsistent with $f_{\text{M}}$ for any $i$. This strongly  constraints  the expected radial velocity curve and  leads to a conservative  upper limit of the NS mass in the system  $\la3$\msun.

These estimates based only on two measurements are highly uncertain but useful for future  determination of the companion and pulsar masses. Our observations showed that the measurements of the companion radial velocity  variation  with the orbital phase with a large telescope  using the medium resolution spectroscopy, as applied in this work, are feasible for this object. 

Such data could  significantly improve our knowledge on the system parameters including reliable   constraints on the companion and NS  masses. Deeper photometry would help to better localise the companion at the CDM and establish its evolution stage. Improved distance measurements would be also useful for this and the other  two companions for which higher resolution spectroscopy is necessary to get more convincing  constraints on their parameters.

\section*{Acknowledgements}

We thank the anonymous referee for useful comments. The work is based on observations made with the Gran Telescopio Canarias (GTC), installed at the Spanish Observatorio del Roque de los Muchachos of the Instituto de Astrof\'isica de Canarias, in the island of La Palma. {\sc iraf} is distributed by the National Optical Astronomy Observatory, which is operated by the Association of Universities for Research in Astronomy (AURA) under a cooperative agreement with the National Science Foundation. Funding for the Sloan Digital Sky Survey IV has been provided by the Alfred P. Sloan Foundation, the U.S. Department of Energy Office of Science, and the Participating Institutions. SDSS-IV acknowledges support and resources from the Center for High Performance Computing at the University of Utah. The SDSS website is www.sdss.org. SDSS-IV is managed by the Astrophysical Research Consortium for the Participating Institutions of the SDSS Collaboration including the Brazilian Participation Group, the Carnegie Institution for Science, Carnegie Mellon University, Center for Astrophysics | Harvard \& Smithsonian, the Chilean Participation Group, the French Participation Group, Instituto de Astrof\'isica de Canarias, The Johns Hopkins University, Kavli Institute for the Physics and Mathematics of the Universe (IPMU) / University of Tokyo, the Korean Participation Group, Lawrence Berkeley National Laboratory, Leibniz Institut f\"ur Astrophysik Potsdam (AIP), Max-Planck-Institut f\"ur Astronomie (MPIA Heidelberg), Max-Planck-Institut f\"ur Astrophysik (MPA Garching), Max-Planck-Institut f\"ur Extraterrestrische Physik (MPE), National Astronomical Observatories of China, New Mexico State University, New York University, University of Notre Dame, Observat\'ario Nacional / MCTI, The Ohio State University, Pennsylvania State University, Shanghai Astronomical Observatory, United Kingdom Participation Group, Universidad Nacional Aut\'onoma de M\'exico, University of Arizona, University of Colorado Boulder, University of Oxford, University of Portsmouth, University of Utah, University of Virginia, University of Washington, University of Wisconsin, Vanderbilt University, and Yale University. The work of AVB, DAZ and AVK was supported by the Russian Science Foundation, grant number 22-22-00921, \url{https://rscf.ru/project/22-22-00921/}. DAZ thanks Pirinem School of Theoretical Physics for hospitality. SVZ acknowledges PAPIIT grant IN119323.

\section*{Data Availability}

The data are available through the GTC archive \url{https://gtc.sdc.cab.inta-csic.es/gtc/}.



\bibliographystyle{mnras}
\bibliography{0621} 

\bsp	
\label{lastpage}
\end{document}
